\documentclass[twocolumn]{aastex631}

\usepackage{natbib}
\usepackage{amsmath}
\graphicspath{{./}{figures/}}

\shorttitle{Shock Breakout w/Atomic Transition lines}
\shortauthors{Morag}

\begin{document}

\title{Radiation Mediated Shock and Planar Shock Breakout in the Presence of Atomic Transition Lines}

\email{jmorag88@gmail.com}

\author[0000-0002-0786-7307]{Jonathan Morag}
\affil{Weizmann Institute of Science \\
234 Herzl St. \\
Rehovot, Israel}



\begin{abstract}

We numerically study fast Newtonian radiation mediated shocks (RMS - $v/c\lesssim0.2$) in two simplified problems within the context of supernova shock breakout; (1) An RMS traveling in a uniform medium, and (2) an RMS escaping a powerlaw density profile in planar geometry ($\rho\propto x^n$).
Both problems were previously solved in the literature assuming a fully ionized plasma medium, emitting and absorbing photons primarily via Bremstrahlung. It was shown that at high shock velocities, Bremstrahlung photon production in the fully ionized scenario is insufficient for achieving local thermal equilibrium (LTE) at the shock front, and as a result the photons can reach Compton equilibrium with electrons temperatures of many keV.

In this study we incorporate, for the first time, opacity from bound species of heavy elements (solar-like composition) into these two problems, at times drastically augmenting the photon production due to the addition of bound-free and bound-bound radiative processes. We solve the problem using a previously developed hydrodynamically coupled multi-group radiative diffusion code, including inelastic Compton scattering and frequency-dependent opacity from the publicly available TOPS table.

We show that the addition of a more realistic opacity leads the radiation to maintain LTE at higher velocities in comparison to the fully ionized problem. In the planar SBO problem this can lead to typically a factor of 2 lower temperature and at times orders of magnitude. This result is important for the observation of supernova shock breakout emission, where strong deviations from LTE and strong X-ray emission were previously predicted for a significant fraction of the parameter space for simple envelope breakout. The SED of SN envelope breakout will very likely remain in LTE for explosions in red super giant stars without stellar wind (and part of blue super giant star explosions), making X-Ray observations less likely in these cases by orders of magnitude relative to previous predictions. We provide a simple semi-analytic description for the SED in the case where LTE is maintained.

A correct calculation of shock-breakout radiation requires opacity tables that include bound yet highly ionized species, ruling out the use of certain line tables (such as the commonly used Kurucz table) for use in the shock breakout problem.

\end{abstract}



\section{Introduction}
\label{sec:intro}

\defcitealias{katz_fast_2010}{KBW10}
\defcitealias{sapir_numeric_2014}{SH14}
\defcitealias{morag_shock_2024}{M24}
\defcitealias{sapir_non-relativistic_2013}{SKWIII}

In core-collapse supernovae (SNe), an explosion in the interior of massive stars launches a radiation mediated shock (RMS) that traverses outwards through the stellar envelope. When the shock reaches the outer edge of the progenitor star or the surrounding material to within an optical depth of $\tau\sim c/v$, a luminous `shock breakout' (SBO) pulse is emitted. In the case of a spherical `envelope breakout' (without surrounding wind or `circumstellar' material) the pulse is predicted to last tens of minutes to an hour, with a luminosity of $\sim10^{45}~\rm erg ~ s^{-1}$, peaking in the UV/X-Ray \citep[see][for a review]{waxman_shock_2016,levinson_physics_2020}. Due to the short time span of the breakout pulse, only a few such SBO candidates have been observed to date \citep{campana_association_2006,soderberg_extremely_2008,gezari_galex_2010,gezari_galex_2015}.

Multiple wavelength breakout observations can shed light on the properties of the progenitor star and its surrounding environment immediately prior to the explosion.
In particular, the extent of wind and circumstellar material present for each SN type is partially undetermined, despite indication of wind interaction in observed lines and X-ray following SBO \citep{smith_mass_2014,bruch_large_2021,dwarkadas_lack_2014,huang_sn_2018,margutti_atel_2013}, and direct observation of a few optically thick ($\tau\gtrsim c/v$) `wind-breakout' events \citep{ofek_interaction-powered_2014,zimmerman_resolving_2023}\footnote{A recent systematic study of optical emission in type II SNe from red super giants suggested that roughly two-thirds of objects do not exhibit wind interaction during the hours and days following SBO \citep{irani_early_2023}. Type Ic/Ibc SNe where the prognitor has likely undergone significant mass loss are associated with X-ray flashes (XRF) and low-luminosity gamma-ray bursts (ll-GRB), indicative of relativistic breakout and/or circumstellar interaction \citep{campana_association_2006,soderberg_extremely_2008,starling_discovery_2011}.}. Upcoming telescope projects should together provide multiwavelength observations of 10's of SBO's in the coming years allowing systematic study, including the Vera Rubin observatory in the optical \citep{ivezic_lsst_2019}, the ULTRASAT UV space mission \citep{shvartzvald_ultrasat_2023}, and the recently launched Einstein probe X-ray mission \citep{yuan_einstein_2022}. UV and X-ray observations are particularly important for determining SBO temperatures. This data should be interpreted with a self-consistent frequency-dependent model of SBO emission, which is currently not available.

The observed spectrum at SBO is determined by the properties of the radiation mediated shock as it reaches the edge of the stellar material.
At high temperatures, the medium is nearly fully ionized, and the dominant photon production mechanism is Bremsstrahlung. Bremstrahlung emission can be insufficient for producing photons at high shock velocities, leading to deviations from local thermal equilibrium (LTE) and high temperatures of many keV. The temperature of RMS and SBO in this case of fully-ionized medium emitting Bermsstrahlung has been solved in the literature \citep{blandford_compton_1981,weaver_structure_1976,katz_fast_2010,sapir_non-relativistic_2013,sapir_numeric_2014}.
Other works study SBO in a partially ionized medium with additional emission/absorption processes, but do not include inelastic Compton scattering. These include works employing the multigroup STELLA radiative transfer code \citep{blinnikov_radiation_2000,tominaga_shock_2011,forster_delay_2018,kozyreva_shock_2020}, which incorporates atomic transition lines based on the Kurucz line list \citep{kurucz_atomic_1995}. The omission of inelastic Compton scattering is reasonable in certain parts of the parameter space
\citep[e.g.][as discussed in \citeauthor{morag_shock_2024} \citeyear{morag_shock_2024}]{kozyreva_shock_2020}, but inappropriate for others, as we discuss below. See also \citet{sapir_numeric_2014} for further comparison of varied SBO temperature results in the literature.


In this work we numerically study the nonrelativistic uniform RMS and planar SBO problems, useful analogs for the envelope SBO problem. In the planar SBO, we study two separate initial density profiles, representative of the profiles in red super giant and blue supergiants stars (RSG,BSG), the likely progenitors of type II and 1987a-like SNe. We include for the first time together in these problems, both inelastic Compton scattering (Kompaneets equation), and frequency-dependent emission and absorption due to radiative photo-ionization (`bound-free') and atomic transition lines (`bound-bound'). We employ a previously developed one-dimensional radiative diffusion code, coupled with hydrodynamics \citep[][hereafter \citetalias{sapir_numeric_2014},\citetalias{morag_shock_2024}]{sapir_numeric_2014,morag_shock_2024}. Radiative processes are handled with a `multi-group' treatment, where the photon flux and energy distribution is binned into frequency groups, and the physics of each group is solved separately. Our code employs the publicly available TOPS opacity table by Los Alamos National Labs (LANL). We verify our results in the case where line opacity is unimportant against results in the literature \citep{katz_fast_2010,sapir_non-relativistic_2013,sapir_numeric_2014}, finding good agreement.

Our results consolidate existing approaches in the literature described above, showing that SBO at lower temperatures and velocities will be affected by the presence of lines and remain near local thermal equilibrium. Then, for faster shocks, behavior will transition towards bulk Comptonization. We will show that for the transition region and at temperatures and velocities below it, the observed temperature is at least a factor of two lower than the predictions of Bulk Comptonization. This reduction of temperature has important consequences regarding the prediction for X-ray flashes in envelope breakout (less so for GRB's), and may eventually provide an additional test for the extent of circumstellar material in RSG's.

We do not address relativistic effects that become important at high velocities and temperatures, such as pair production \citep{budnik_relativistic_2010,nakar_relativistic_2012,ito_monte_2020,ito_monte_2020-1}. We also do not examine the wind-breakout case, where the possible presence of collisionless shocks can yield a strong deviation from LTE behavior and many keV temperatures \citep{waxman_tev_2001,katz_x-rays_2011,murase_probing_2014,wasserman_optical_2025}.

This paper is structured as follows. In \S~\ref{sec: Numeric Description} we describe our numerical code, including details of the frequency-dependent opacity. In \S~\ref{sec: URMS intro} we study the uniform RMS problem, including describing initial conditions in \S~\ref{sec: U-RMS init cond} and the temperature results in \S~\ref{sec: U-RMS T results}. In \S~\ref{sec: pSBO} we perform a similar analysis for the planar shock breakout, with temperature results in \S~\ref{sec: pSBO results}. In \S~\ref{sec: Doppler} we address the finite frequency resolution in our code relative to the line resolution. In \S~\ref{sec: Observations} we discuss the impact of our theoretical results on upcoming observations, particularly in the X-ray bands. We discuss and summarize in \S~\ref{sec: summary}.  

\section{Numerical Details - Hydrodynamics equations}
\label{sec: Numeric Description}

In both the uniform RMS and the planar SBO problems we use a numerical code previously described in \citetalias{morag_shock_2024} and in \citetalias{sapir_numeric_2014}. Our code describes the evolution of Lagrangian matter density $\rho$, velocity $v$, matter pressure $p$, and radiation energy and flux $u_\nu$ and $j_\nu$. $u_\nu$ and $j_\nu$ are calculated using a `multigroup' treatment, where the photons are binned into energy groups and the radiation physics is solved separately for each group. In all simulations we include compression of the radiation and plasma, radiative diffusion, radiative emission and absorption, as well as inelastic Compton scattering via the Kompaneetz equation.

The governing equations of our code describing the evolution with the above processes are given by equations 24-36 in \citetalias{morag_shock_2024}. We make two small substitutions in these equations to account for the planar geometry (as opposed to spherical geometry in \citetalias{morag_shock_2024}). First, we substitute $\nabla\cdot f=\frac{1}{r^2}\frac{\partial ( r^2 f)}{\partial r} \to \frac{\partial f}{\partial r}$ in their equations 30-32, 36, where $f$ is an arbitrary function. Second, we change their equation 25 to
\begin{equation}
    \rho = \rho_{\rm init} \frac{\partial r_{\rm init}}{\partial r},
\end{equation}
where $\rho_{\rm init}$ and $r_{\rm init}$ are the initial position and density (denoted with a subscript $0$ in \citetalias{morag_shock_2024}). The rest of the equations are used in our code as written.


We produce two types of simulations with our code, each employing a different choice of frequency-dependent opacity. In one type (henceforth denoted `free-free only'), we choose that the emission and absorption opacity $\kappa_{\rm abs,\nu}$ only includes Bremstrahlung  (given analytically in \citetalias{sapir_numeric_2014} equations 7-9). In the other set of simulations we use a coarse multi-group frequency average of $\kappa_{\rm abs,\nu}$, the high-frequency resolution TOPS opacity table. As is done in \citetalias{morag_shock_2024}, we differentiate in our governing equations between the frequency-dependent diffusion opacity $\kappa_*$, employed in \citetalias{morag_shock_2024} equations 26 and 33, and the frequency dependent emission / absorption opacity $\kappa_{\rm abs}$ used in their equation 29. The diffusion opacity is given by, $\kappa_*^{-1}= (\Delta\nu_{\rm bin})^{-1}\int d\nu(\kappa_{\rm es}+\kappa_{\rm abs,\nu})^{-1}$, where $\kappa_{\rm es}$ is the Thomson scattering opacity and $\Delta\nu_{\rm bin}$ is the width of the binned frequency group in the simulation. Meanwhile, the absorption opacity of each group is given by $\kappa_{\rm abs}= (\Delta\nu_{\rm bin})^{-1}\int d\nu\kappa_{\rm abs,\nu}$. In the free-free only simulations, analytic values sampled at each group are used both in $\kappa_*$ and in $\kappa_{\rm abs}$, without performing an average.

The numerical scheme and the time-step limitations are the same as in \citetalias{morag_shock_2024} sec. 3.2. In TOPS simulations we include the `protections' on diffusive and radiative transfer, including the addition of a flux limiter, previously described in \citetalias{morag_shock_2024} sec. 3.2.2 ( their equations 40-41).

We assume everywhere a solar composition of elements, and use a slightly simpler equation of state for the matter energy and pressure, which does not include latent energy of ionization, without affecting our conclusions. Namely, the matter energy density, temperature and pressure are related by
\begin{equation}
    e=\left(\frac{1}{\gamma-1} \right)p, \, \, p=(\rho/\mu) T, \, \, \mu=m_{\rm amu} / \sum_i (1+Z_i) x_i/A_i
\label{eq:EOS}
\end{equation} 
where $\gamma$ is adiabatic index, approximated as 5/3, $m_{\rm amu}$ is the atomic mass unit, $x_i$, and $A_i$ are the mass fraction and atomic mass of element $i$, $Z_i$ is the atomic number.

 
For later use we define the equivalent LTE temperature $T_{\rm LTE}$ as 
\begin{equation}
    T_{\rm LTE}= ( u / a_{\rm BB} )^{1/4},
    \label{eq: T_LTE def}
\end{equation}
where $u$ is the frequency integrated photon energy density, and $a_{\rm BB}$ is the Boltzmann blackbody constant. We also introduce an approximate photon temperature defined by the photon energy distribution $u_
\nu$ \citep{zeldovich_physics_2002}.
\begin{equation}
    T_{\gamma}=\frac{1}{4u}\int_{0}^{\infty}\left[h\nu u_{\nu}+\frac{c^{3}}{8\pi}\left(\frac{u_{\nu}}{\nu}\right)^{2}\right]d\nu
    \label{eq: T_gamma}
\end{equation}
The latter two temperatures are equivalent when the radiation is in LTE, and $T_\gamma$ matches the plasma temperature $T_{\rm e}$ when the two are either in LTE or in Compton equilibrium.

\subsection{The Opacity Table}
\label{sec: The Opac Tbl}
The TOPS opacity table \citep{colgan_new_2018} is to our knowledge the only publicly available source for frequency-dependent opacity that includes highly ionized species. The table is experimentally verified, though at densities much higher \citep[$\sim 10^{-2}-10^{-3} \rm ~g~ cm^{-3}$][]{colgan_light_2015,colgan_new_2016,colgan_new_2018} than we solve for in our problem ($10^{-11}-10^{-7} \rm ~g~ cm^{-3}$). Thus, the existence of individual lines or atomic states used in TOPS is not assured, though we are concerned here with the ensemble behavior describing the general SED, and delayed deviations from LTE due to the cumulative presence of atomic lines. The table assumes thermal equilibrium between the plasma and the radiation, and does not include `expansion opacity effects' (see \S ~ \ref{sec: summary}).

Unlike our previous use of the table in \citet{morag_shock_2023,morag_shock_2024}, our downloaded TOPS Opacity table contains all elements up to and including Zn (with the exception of Li, Be, and B). Adding heavy elements, despite their very low abundance, turned out to be important at high (10's-100's eV) temperatures due to the admittance of highly ionized (but not fully ionized) states and their associated lines. We also note that the TOPS table we downloaded is capped at $h\nu=10^4$ eV. In cases where higher frequencies were needed, we complemented the table with analytic Bremstrahllung-only opacity at $h\nu>10^4$ eV. TOPS is limited in resolution with regards to the underlying temperature, density, and frequency grid. This limitation can lead to crashes in numerics and impacts our ability to test convergence, especially since the frequency resolution is at least an order of magnitude coarser than the width of individual lines  (see \S ~ \ref{sec: summary}).

\section{Uniform Radiation Mediated Shock}
\subsection{U-RMS Introduction}
\label{sec: URMS intro}

In the steady-state non-relativistic uniform radiation mediated shock (RMS) problem, a strong shock traverses a uniform plasma medium at velocity $\beta_0=v_0/c$ in planar geometry. The medium is assumed to be highly ionized or fully ionized. Energy and momentum are dominated by photons undergoing diffusion with scattering opacity $\kappa_{\rm es}\sim \sigma_T/m_{\rm i}$, where $\sigma_T$ is the Thomson cross section, and $m_{\rm i}$ is the mass of the dominant ion in the plasma (e.g. the proton mass for Hydrogen).

Hydrodynamics for the fully-ionized RMS was solved by \citet{weaver_structure_1976}.
In the shock frame, cold `upstream' plasma heading towards the shock at velocity $v_0$, collides with hot plasma at slower velocity, $v_0\left( \frac{\gamma-1}{\gamma+1}\right)=v_0/7$, and gets swept `downstream'.
The upstream protons and ions transfer their energy and momentum to the photons across a length scale $l\sim (\rho \kappa_{\rm es} \beta_0)^{-1}$, which determines the thickness of the shock. This scale, across which the plasma decelerates, also coincides with the distance to which photons diffuse upstream against the flow. Correspondingly, the scattering optical depth across the shock is $\tau \sim l (\rho \kappa)^{-1} \sim c/v$ and the shock-crossing time-scale is $t_0 \sim c/\rho \kappa v_0^2$ \citep{katz_fast_2010}.

As the plasma heats up across the shock, it emits (and absorbs) photons. The hydrodynamics and total energy density are insensitive to the exact photon production mechanism, but the radiative energy density distribution $u_\nu$ and the plasma temperature\footnote{The plasma temperature $T$, represents the temperature of both the electrons and the ions, which are assumed to be in thermal equilibrium throughout, a reasonable assumption in this scenario.} $T$ can be greatly effected. If emission (and absorption) processes are efficient relative to the shock crossing timescale, the radiation will maintain a Planck distribution $B_\nu(T)$ throughout. In such a scenario, the local temperature will everywhere be given by
\begin{equation}
    T_{\rm LTE,W}=[3 P_{\rm W} /a_{BB}]^{1/4}
    \label{eq: T LTE U-RMS}
\end{equation}
where $a_{BB}=8\pi^5/15(hc)^3$ and $P_{\rm W}(r,t)$ is the pressure, as described by the Weaver steady-state analytic solution \citep{weaver_structure_1976,katz_fast_2010,sapir_numeric_2014}. 

For Bremsstahlung emission, the opacity is highest at lowest frequencies, and behaves like  $\kappa_\nu\propto\nu^{-2}$. At faster velocities where the emission (and absorption) are inefficient, there exists a critical frequency $h\nu_{\rm c}$ \citep[][hereafter \citetalias{katz_fast_2010}]{katz_fast_2010}, above which emitted photons will undergo significant inelastic Compton scattering with the hot electrons before being reabsorbed again\footnote{Efficient Compton scattering requires that the $y$ parameter at the shock, $y=(4T/m_ec^2)\tau_{\rm es}^2>1$, which generally holds for the relevant part of the parameter space.}. Photons below $h\nu_{\rm c}$ will thermalize with the electrons, forming the Rayleigh Jeans tail of a Planck distribution $B_{\nu}(T)$. Photons above $h\nu_{\rm c}$ will up-scatter to $\sim3T$ energies, forming a Wien peak. $T$ in turn -assuming Comptonized photons dominate the energy density- is determined by the equation of state $T=p_\gamma/n_\gamma$, where $p_{\gamma},n_{\gamma}$ are the photon pressure and number density of Comptonized photons. Since $p_\gamma=p_{\rm W}$ depends only the hydrodynamics and is insensitive to photon production, the temperature in the up-scattered Wien peak is determined inversely by the photon production rate $Q_{\rm eff}$. The $Q_{\rm eff}$ is integrated in frequency from all photons emitted at $h\nu>h\nu_{\rm c}$ and for the case of Bremstrahlung opacity only, is given by \citepalias{katz_fast_2010}
\begin{equation}
    Q_{\rm eff}\approx\alpha_{\rm e} n_{\rm i} n_{\rm e} \sigma_T c \sqrt{\frac{m_{\rm e}c^2}{T}}\Lambda_{\rm eff}g_{\rm eff},
    \label{eq: Q_eff approx}
\end{equation}
where $n_{\rm i}$ and $n_{\rm e}$ are the number densities of ions and free electrons, and $\sigma_T$ is the Thomson cross-section, $m_{\rm e}$ is the electron mass and $c$ is the speed of light. The coefficient $\Lambda_{\rm eff}\sim\log(\lambda)$ and the effective gaunt factor $g_{\rm eff}(\lambda)$, both result from integrating the emission from $h\nu_{\rm c}$ to $+\infty$ and are a weak function of $\lambda=T/h\nu_{\rm c}$.
We note that the description of $Q_{\rm eff}$ for $\dot{n}_\gamma$ is approximate since it assumes that all the photons are Wien distributed and that photon absorption for $h\nu>h\nu_{\rm c}$ is negligible. We show below that these approximations are reasonable for a large part of our parameter space.

\citetalias{katz_fast_2010} provide analytic descriptions for the deviations in temperature from LTE in steady state. From peak Compton temperature $T_{\rm peak}$ near the shock, photon production will reduce the temperature towards $T_{\rm LTE}$ in the far downstream across a thermalization length scale $L_T \sim v_0 n_{\gamma,\rm peak}/Q_\gamma$ (assuming uniform photon production rate $Q_{\rm eff}(\tau>0)=Q_{\rm eff}(T_{\rm peak})$, where $\tau>0$ represents the downstream optical depth). Therefore, by comparing $L_{\rm T}$ to the shock width $l$, we arrive at a velocity cutoff $\beta_{0,\rm cut}$ above which $T$ will deviate from $T_{\rm LTE}$ to higher temperatures \citepalias{katz_fast_2010}. Namely,
\begin{equation}
    \beta_{0,\rm cut} > 0.7 n_{\rm p,15}^{1/30} (\Lambda_{\rm eff} g_{\rm eff})^{4/15}, 
\end{equation}
where $n_p=n_{\rm p,15}10^{15} \rm g ~ cm^{-3}$ is the proton density. Likewise, we may estimate a relationship between the peak temperature and the shock velocity, given by
\begin{equation}
    \beta_{0,K10} \approx 0.2 \left( \Lambda_{\rm eff} g_{\rm eff}/ 2 \right)^{1/4} (T_{\rm peak}/10 \, \rm keV),
    \label{eq: katz_Tpeak}
\end{equation}
where $\Lambda_{\rm eff}(\lambda)$ and $g_{\rm eff}(\lambda)$ can be solved for self-consistently. These results agree with \citet{weaver_structure_1976} for $T<50keV$. 

\citetalias{katz_fast_2010} also provide an ODE description that solves for the number density of photons $n_\gamma(\tau)$ in steady-state under similar assumptions to the heuristic analytic solutions, and arrive at similar results. Here we rewrote these relations for the case of a general plasma composition, in lieu of a Hydrogen only composition as was done in \citetalias{katz_fast_2010}. We note that for the Hydrogen-dominated (but not Hydrogen-only) envelope that we tested here, the result for the Bremsstrahlung only case is nearly unaffected by the presence of other elements.

    
\subsection{Initial and Boundary conditions}
\label{sec: U-RMS init cond}

The calculation describes a constant velocity piston being driven into a cold medium of uniform density $\rho_0$, driving a radiation mediated shock with velocity $v_0=\beta_0 c$\footnote{Corresponding piston velocity in the unshocked plasma frame is $(6/7)v_0$}. For faster convergence, the simulation is started with the analytic RMS solution \citep[][Appendix C]{weaver_structure_1976,sapir_numeric_2014}. After several shock crossings, whose number depends on the opacity and on the velocity, the shock reaches steady state, both in hydrodynamics and in temperature. Higher shock velocities in the free-free only case require many shock crossings, up to $\sim 150$ in order to reach steady state. The system is built large enough so that the shock will not reach the edge prior to arriving in steady state. The initial plasma temperature is $T_{\rm LTE,W}$ (equation \ref{eq: T LTE U-RMS}), and the corresponding radiation distribution is given $u_\nu=4\pi B_\nu(T_{\rm LTE,W})$/c, where $B_\nu$ is the Planck intensity.

We use a constant initial spatial grid. Simulations vary in grid spacing between 15 and 60 spatial grid points per shock crossing and a multigroup frequency grid of $\Delta \nu/\nu \sim 0.07 - 0.25$. At steady state, all sims are converged to better than a percent in $\rho,v,p$, as well $3\%$ in peak T and $10's$ of percent in $u_\nu$ at individual frequencies.

Convergence of the temperature peak in the ff-only case, requires very low initial temperatures in the upstream due to the strong dependence of photon count on temperature (for numerical reasons the upstream temperature cannot be set identical to zero). When using TOPS instead, the highest possible temperature for the cold matter is taken, $T_{\rm init}>1eV$, while ensuring that $u_{\rm ph,\,upstream}\ll u_{\rm ph,\,downstream}$ and that the results are not effected. Inelastic Compton scattering and bound-free and bound-bound opacity (where relevant) are turned on gradually, over roughly a shock crossing time to avoid numerical issues. 

\subsection{U-RMS Temperature Results}
\label{sec: U-RMS T results}

\begin{figure}
    \centering
    \includegraphics[width=\columnwidth]{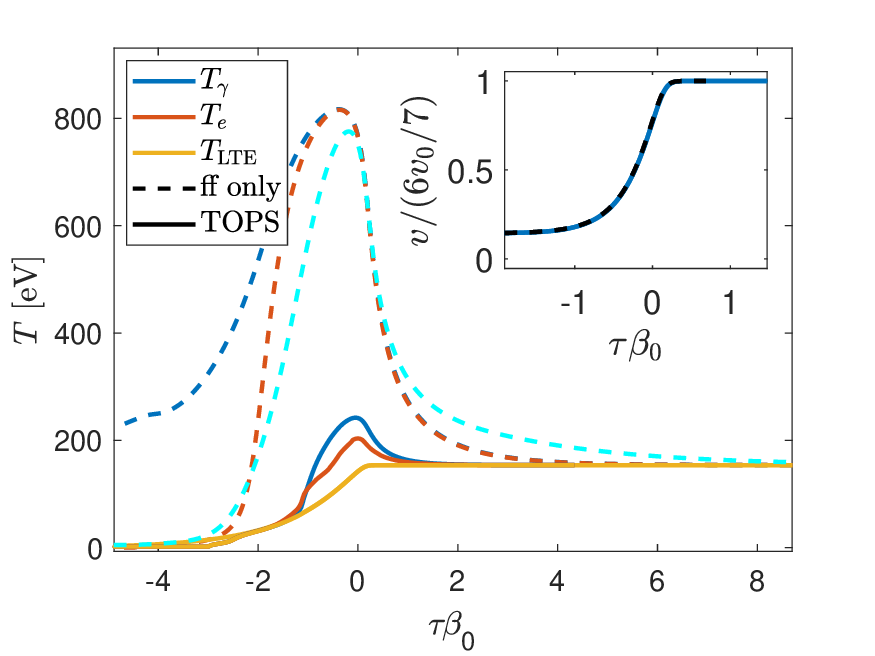}
    \caption{Temperature profile of a steady-state RMS as a function of optical depth, for a shock velocity $\beta_0=0.1$ and initial density $\rho_{0}=10^{-9} \rm \, g \, cm^{-3}$. Photon temperature  $T_\gamma$ (equation \ref{eq: T_gamma} - in blue) and plasma temperature $T$ (in red) are markedly lower in the presence of TOPS opacity (solid lines) relative to the ff-only case (dashed lines). $T_{\rm LTE}$ (equation \ref{eq: T_LTE def} in yellow) is nearly identical for both cases. ff-only simulation results agree with the result of an ODE integration, described in the appendix of \citetalias{katz_fast_2010} (dashed teal lines).  The dashed ff-only results agree with figure 1 in \citetalias{sapir_numeric_2014}. Inset shows simulation velocity in the shock frame for both the ff-only and TOPS case (both in blue), against the analytic Weaver solution (black dashed), all collapsed on top of each other. $\tau=0$ is set to the point of maximum convergence of the flow and positive $\tau$ indicates downstream. $\tau \beta_0=1$ corresponds to one shock crossing width.
    }
    \label{fig: U-RMS T profile}
\end{figure}

In this section we present numeric results showing the effect of opacity on the uniform RMS problem. For each choice of parameters, we produce two sets of simulations, with and without the effect of bound-free and bound-bound processes in the opacity. This change affects the plasma temperature $T$ and photon energy distribution $u_\nu$, but has negligible effect on the radiation pressure $p_\gamma \propto \int u_\nu d\nu$, which dominates the total pressure $p$. Converged simulation results agree in ($\rho$, $v$, $p$) with the analytic Weaver solution \citep[][Appendix C]{weaver_structure_1976,sapir_numeric_2014} to 2\% or better\footnote{In the cases $\beta=0.2-0.25$, $\rho_0=10^{-7} \, \rm g \, cm^{-3}$ with free-free opacity only, pressure agrees with the analytic solution to 3\%.}. In the inset to figure \ref{fig: U-RMS T profile}, we show an example of a steady-state velocity profile in a simulation compared with the analytic RMS solution \citep{weaver_structure_1976,sapir_numeric_2014}.

\begin{figure}
    \centering
    \includegraphics[width=\columnwidth]{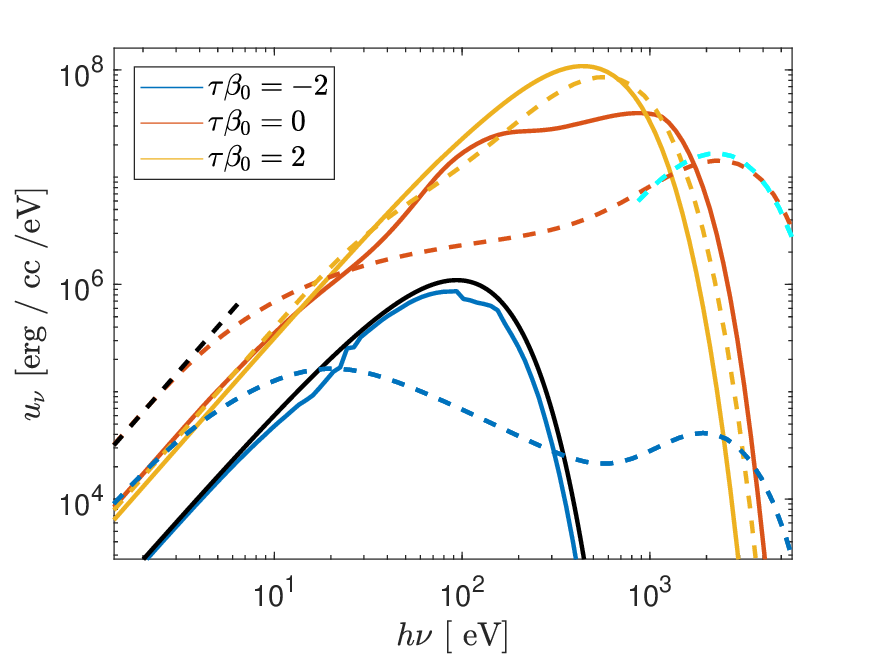}
    \caption{Steady-state photon energy density distribution, $u_\nu$ for the same simulations as in figure \ref{fig: U-RMS T profile}, showing that adding TOPS opacity reduces $T_\gamma$ and brings the distribution closer to LTE. In the presence of ff-only opacity (dashed lines) $u_\nu$ reaches very high temperatures, matching Compton equillibrium (teal dashed line) based on $T_\gamma$ from figure \ref{fig: U-RMS T profile}.
    $u_\nu$ cross-sections are shown at (-2,0,2) shock crossings from the point of max compression (with positive $\tau$ indicating downstream). }
    \label{fig: U-RMS u_nu example}
\end{figure}

Figures \ref{fig: U-RMS T profile} and \ref{fig: U-RMS u_nu example} show an example with parameters  ($\beta_0=0.1$, $\rho_0=10^{-9} \, \rm g\, cm^{-3}$). We first examine the result in presence of free-free opacity only. Figure \ref{fig: U-RMS T profile} provides a snapshot of the steady-state temperature across the shock. In the case where only free-free opacity is present (dashed lines), plasma and photons temperatures $T$ and $T_{\gamma}$ at the shock and downstream of it are nearly matched in Compton equilibrium. The temperature deviates largely from $T_{\rm LTE}$, reaching peak temperatures near the shock of $\sim 800$ eV before settling into equillibrium ($T_{\rm LTE} \sim 160$ eV) in the far downstream. Our free-free only results are in close agreement with the results of \citetalias{sapir_numeric_2014}, figure 1 ($\beta=0.1$, number of protons $n_{p}=\rho_0/m_{\rm p}=10^{17} \, cm^{-3}$). In \citetalias{sapir_numeric_2014}  they solve a hydrogen only problem, and get similar results, since Bremstrahllung opacity is insensitive to metallicity. We also observe a low energy density radiative precursor (in agreement with \citetalias{sapir_numeric_2014} - see their discussion).

Figure \ref{fig: U-RMS u_nu example} shows cross sections of the radiation energy density distribution $u_\nu$, in the vicinity of the shock. In the case of free-free opacity only (again in dashed lines), $u_\nu$ at the shock approaches Compton equilibrium with a corresponding Wien peak (red dashed line). Near downstream of the shock (yellow dashed line), $u_\nu$ approaches LTE. Meanwhile, in the near upstream, diffusion of photons from the shock leads $u_\nu$ to exhibit a similar (though  less prominent) high-frequency peak (blue dashed line). The latter is the high temperature, but relatively low energy radiative precursor.

Returning to figure \ref{fig: U-RMS T profile}, we compare the temperature result for $\beta = 0.1$ (free-free only) to the semi-analytic ODE estimate presented in \citetalias{katz_fast_2010} (dashed teal line), which assumes Compton equilibrium with an equation of state $P=n_\gamma T$, where $n_\gamma$ is the number density of photons. We find good agreement between the ODE and the sims at the shock and immediately downstream of it,  where the radiation in the simulation is near Compton equilibrium. Peak temperature in the ODE is in general up to 30\% lower than the temperature observed in the simulation, primarily since a non-negligible a fraction of the energy and photons lie outside of the Wien peak at lower frequencies (e.g. red dashed line in figure \ref{fig: U-RMS u_nu example}). $Q_{\rm eff}$ in the ODE also agrees with $\dot{n}$ from the simulation to 10's \% or better. The far downstream temperature in the ODE progresses towards $T_{\rm LTE}$, due to a heuristic suppression factor on the photon production, leading the net photon production to be zero when $T_{\rm ODE}=T_{\rm LTE}$. The details of our implementation of the ODE are presented in the appendix.

In figure \ref{fig: U-RMS u_nu example} again, we compare the result at $\tau=0$ to a Wien spectrum using $(T,u)$ derived from the semi-analytic ODE estimate (teal dashed line -  \citetalias{katz_fast_2010}). We find very good agreement at the Wien peak. We compare the low-frequency end at the location with a Planck energy density $B_\nu(T)$, using the same $T$ from the ODE. We again find excellent agreement, since the low frequency photons are in thermal equilibrium with the plasma due to the large Bremstrahllung opacity $\kappa_\nu$.

When we introduce the bound-free and bound-bound opacity using TOPS (solid lines in figures \ref{fig: U-RMS T profile} and \ref{fig: U-RMS u_nu example}) the temperature is drastically different. Increased photon production reduces the temperature everywhere towards $T_{LTE}$, with minor deviations from LTE located at the peak (similar location to the previous peak). When the temperature does deviate from LTE, $T$ and $T_\gamma$ are not matched as in the free-free only case, since the spectrum is not dominated by a single Wien peak. The high-temperature radiative precursor disappears, as photons diffusing upstream are reprocessed.

\begin{figure}
    \centering
    \includegraphics[width=\columnwidth]{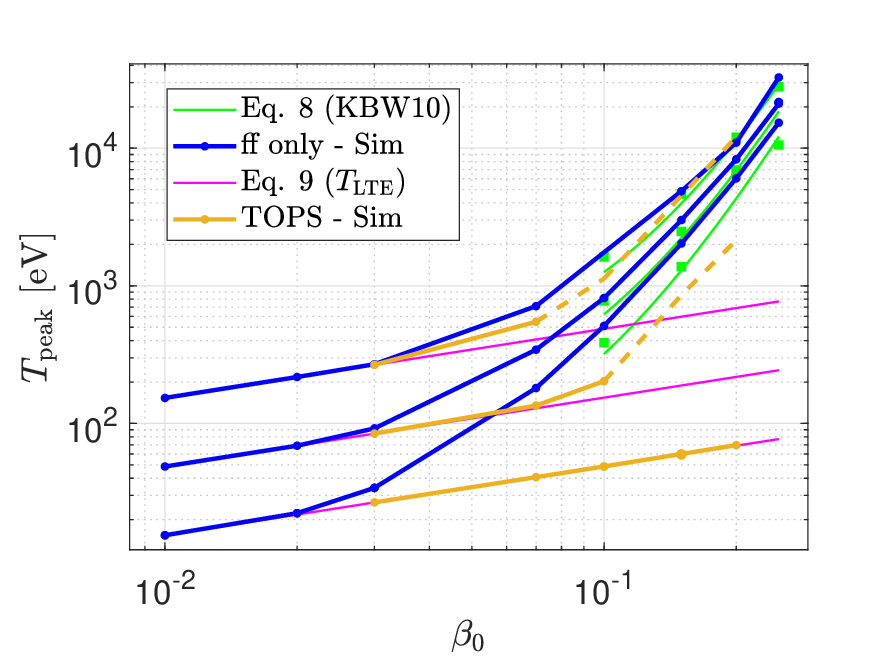}
    \caption{Mapping of peak electron temperature of as function of $\beta_0$, for three different densities. From top to bottom, the initial densities presented are $\rho_0=10^{-7},10^{-9},10^{-11} \, \rm g \, cm^{-3}$. For high velocities, calculations with free-free only opacity increase, in agreement with equation \ref{eq: katz_Tpeak}. The presence of TOPS opacity can significantly delay deviation from $T_{\rm LTE}$, especially at lower densities.  Every dot in the figure represents a converged calculation}
    \label{fig: U-RMS mapping}
\end{figure}

In figure \ref{fig: U-RMS mapping}, we map peak electron temperature as a function of density and shock velocity for the cases of free-free only opacity and full opacity (blue dots and connecting line and yellow dots and connecting line). At the lowest shock velocities, temperature for both opacity choices remains  remains close to (magenta line)
    \begin{equation}
    T_{\rm LTE,p } = ( 6 \, (3 \rho_{0} v_{0}^2)\,/ 7a_{\rm BB})^{1/4}.
    \label{eq: T LTE RMS peak}
\end{equation}
As shock velocities increase past $\beta\sim0.03$ in the free-free only case, peak temperature gradually increases away from $T_{\rm LTE}$ and follows equation \ref{eq: katz_Tpeak} (green line) to 10's of percent. Equation \ref{eq: katz_Tpeak} itself is in close (few percent) agreement with individual calculations of the ODE presented in \citetalias{katz_fast_2010} (green dots), based on similar assumptions.

\begin{figure}
    \centering
    \includegraphics[width=\columnwidth]{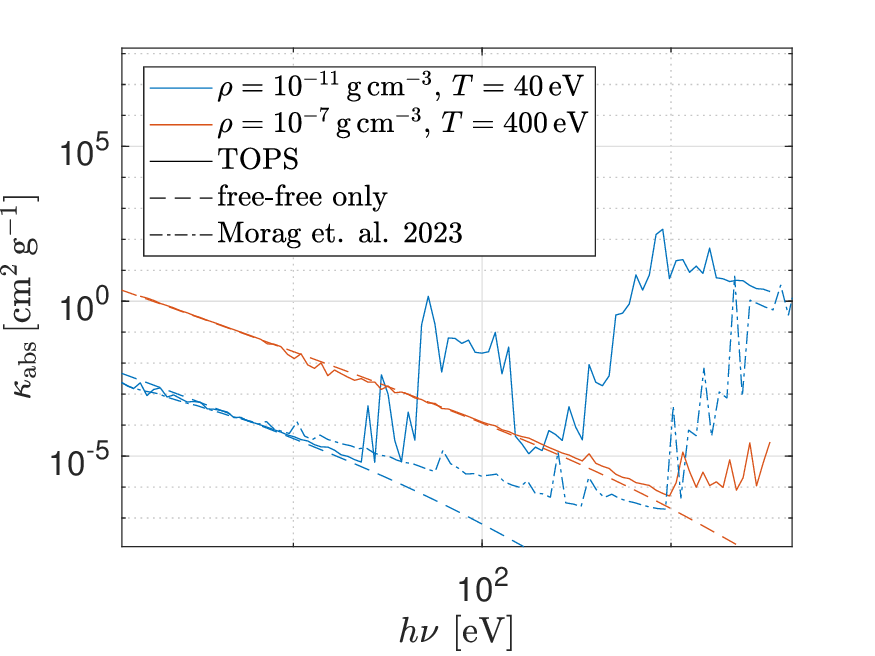}
    \caption{Example multigroup emission / absorption opacities at $T=(40,400)$ eV, which correspond to $T_{\rm LTE}$ at velocity $\beta_0=0.07$ and densities $\rho=(10^{-7},10^{-11}) \, \rm g \, cm^{-3}$ (see figure \ref{fig: U-RMS mapping}). The effect of TOPS opacity on photon production relative to free-free only opacity is pronounced at lower breakout temperatures, $T=40$ eV, but has only a minor effect at $T=400$ eV. The table from \citet{morag_frequency_2023} is also shown at 40 eV, demonstrating a drastically different result due to use of the Kurucz table. Photon production drops roughly exponentially above the Planck peak frequency, located here at $h\nu\sim(100,1000)$ eV.}
    \label{fig: knu_avg_example}
\end{figure}

The introduction of the TOPS opacity in figure \ref{fig: U-RMS mapping} (yellow line) delays deviations from $T_{\rm LTE}$ in many parts of the parameter space. The relative effect of TOPS is mostly determined by temperature, to which the opacity is most sensitive. At highest densities ($\rho_0=10^{-7} \, \rm g \, cm^{-3}$), and therefore highest correspong temperatures $T_{\rm LTE}\propto (\rho_0 \beta_0^2 c^2)^{1/4}$, all elements are nearly fully ionized, and the TOPS opacity is similar to a Bremstrahllung-only opacity (example in figure \ref{fig: knu_avg_example} - red lines). Consequently, the deviation from the free-free only opacity in this case is minor. Meanwhile, at lower densities, the breakout temperature is lower, and the opacity is markedly different with TOPS (figure \ref{fig: knu_avg_example} - blue lines). In the case of $\rho_0=10^{-11} \, \rm g\, cm^{-3}$, simulations with TOPS remain in LTE for the entire parameter space that we checked, deviating by orders of magnitude from the temperature prediction in the presence of free-free only opacity. Figure \ref{fig: knu_avg_example} also includes \citet{morag_frequency_2023} opacity tables for comparison (see discussion).

\section{Shock Breakout in Planar Geometry}
\label{sec: pSBO}
\subsection{Planar SBO Introduction}
\label{sec: pSBO Introduction}
The problem of a shockwave escaping a powerlaw initial density profile $\rho_{\rm init} \sim x^{n}$, also known as the `planar shock breakout problem', was introduced and solved by \citet{sakurai_problem_1960} and then later by \citet{gandelman_shock_1956}. The shock structure and emitted luminosity for the case of a diffusive shock was solved by \citet{sapir_non-relativistic_2011}, by combining in an Anzats the Sakurai solution with the analytic Weaver RMS. The emitted temperature was calculated by  \citet{sapir_non-relativistic_2013} and the multi-group output spectrum in \citetalias{sapir_numeric_2014}, both assuming free-free emission.

As the shock travels towards the stellar edge ($x\to0$), its velocity is given by
\begin{equation}
    v_{\rm sh} \sim x^{-\beta_1 n},
    \label{eq: vsh Sakurai}
\end{equation}
where $\beta_1\approx0.18$ for common choices of $n=3/2,3$, corresponding to  convective and radiative stellar envelopes, respectively. Its hydrodynamic profile follows the Sakurai-Weaver Anzats. Breakout occurs when the optical depth ahead of the shock is $\tau=c/v_{\rm bo}$. The breakout velocity $v_{\rm bo}$ and density $\rho_{\rm bo}$ are defined self-consistently at this time, where $\rho_{\rm bo}$ is the initial unshocked density profile at the shock location. At breakout the corresponding length of the shock is $l \sim c / \kappa \rho_{\rm bo} v_{\rm bo}$.
The time scale for this breakout is $t_{\rm bo}= c / \kappa \rho v^2$, based on the shock crossing time scale. Following breakout, shock-heated material is expelled in the direction of the original shock.

The emitted bolometric luminosity for planar shock breakout was given semi-analytically by \citet{sapir_non-relativistic_2011} as a function of $\rho_{\rm bo}$, $v_{\rm bo}$. The peak luminosity per unit area is approximately given by $\mathcal{L}_{\rm peak} \approx \rho_{\rm bo}v_{\rm bo}^3$ corresponding to the shock's energy.

Similarly to the previous discussion in the uniform RMS problem, opacity is dominated by free-electron scattering. The hydrodynamics ($\rho,v,p$) and bolometric breakout luminosity are insensitive to the photon production mechanism, though temperature $T$ and radiation distribution $u_\nu(r,t)$ and correspondingly the breakout spectrum, can be greatly affected. In cases where the photon production is efficient within a shock crossing time, the radiation will match a Planck spectrum $u_\nu=B_\nu(T)$. The temperature everywhere in this case is very nearly given by
\begin{equation}
    T_{\rm LTE,SW}=[3 P_{\rm SW} /a_{BB}]^{1/4},
    \label{eq: T LTE S-W}
\end{equation}
where $P_{\rm SW}(r,t)$ is the pressure in the Sakurai-Weaver Anzats \citep[][different from $P_{\rm S}$ in equation \ref{eq: T LTE U-RMS}]{sapir_non-relativistic_2011}, which describes well the pressure in the simulation.

For fully ionized Bremsstrahlung emission, \citet{sapir_non-relativistic_2013} provide an analytic formula for the planar shock breakout peak emission temperature $\mathcal{T}_{\rm peak}$ and  a semi-analytic description for the time-dependent emission temperature $\mathcal{T}(t)$, given by
\begin{align}
    \log_{10} & \left( \frac{\mathcal{T}_{\rm peak}}{10 ~ \rm eV} \right) = 1 + 1.69 \, \beta_{-1}^{1/2} \notag \\
    &+ \left(0.26 - 0.08 \, \beta_{-1}^{1/2}\right) \log_{10}\left(A n_{\rm p,15}\right), \quad n=3 \notag \\
    \log_{10} & \left( \frac{\mathcal{T}_{\rm peak}}{10 ~ \rm eV} \right) = 0.95 + 1.78 \, \beta_{-1}^{1/2} \notag \\
    & + \left(0.26 - 0.08 \, \beta_{-1}^{1/2}\right) \log_{10} \left(A n_{\rm p,15}\right), \quad n=3/2,
    \label{eq: T_peak SKWIII}
\end{align}
where $\beta_{\rm bo}=0.1\beta_{-1}$, $A$ is the ion mass in atomic mass units, and $n_p=10^{15}n_{p,15}$. The formulas were calibrated against a set of gray simulations (with 10\% agreement)  counting total photon production rate $\dot{n}_\gamma(r,t)$ and assuming Compton equilibrium. They found that peak temperature occurs roughly one shock-crossing time prior to peak luminosity time.

\subsection{Initial and Boundary Conditions}

Our numeric simulation describes an RMS shock traversing in a cold medium, with initial matter density $\rho_{\rm init} = \rho_{\rm bo}\left( \beta_{\rm bo}\tau \right)^{n/(n+1)}$, where $x=0$ and $\tau=0$ represents the stellar edge. We introduce the terms `inner' and `outer' edges to denote densest and least dense simulated edges of the problem (respectively), in analogy to the geometry of a stellar explosion. For faster convergence, initial shock hydrodynamics are started using the Sakurai-Weaver approximation \cite{sapir_non-relativistic_2011}. The inner piston at the high density edge is taken to be the \citet{sakurai_problem_1960} solution.
The photons and the plasma are initially assumed to be in LTE (Planck distributed in agreement with a local plasma temperature $T$).

Similarly to the initial spatial grid in \citet{morag_shock_2023} and \citetalias{morag_shock_2024}, our spatial resolution is a smooth function of position, described consecutively from the piston to the outer edge as follows. We place modest resolution near the piston, highest resolution at the initial shock location, gradually decreasing resolution towards the edge (for constant cell number across the progressing RMS shock), flat resolution for $\tau \lesssim c/v$, and increasing resolution towards the outer edge  ($\tau<10^{-2}-10^{-3}$).
Simulations are started between 4-8 shock crossing distances away from the stellar edge, and are shown to be largely insensitive to the choice of distance.
We employ a multigroup frequency grid of $\Delta \nu/\nu \sim 0.07 - 0.25$. Unless otherwise indicated, all sims are converged to better than 2-3\% in $\rho,v,p$, as well $3\%$ in peak T and $10's$ of percent in emitted spectral energy distribution (SED) $L_\nu$ at individual frequencies. We also check convergence with respect to initial pre-shock temperature, minimum optical depth in the outermost cell ($\tau\sim10^{-2}-10^{-3}$), and location of the piston relative to the starting shock location. 

To avoid numeric issues associated with the presence of strong lines at low temperatures, in some of the SBO simulations we reduce the multigroup opacity table for the range $h\nu>\chi\equiv200$ eV to its value at $h\nu=\chi$, only if the former is higher. Namely we make the substitution, $\kappa_{\rm abs,\nu}(h\nu>\chi)\to \min[\kappa_{\rm abs,\nu}(h\nu=\chi),\kappa_{\rm abs,\nu}(h\nu>\chi)]$. 
Reducing peaks in this way does not negate our conclusions as long as the photons remain in LTE, since we are only reducing $\kappa$ (photon production) and never increasing it. This substitution is made for all simulations with $\rho_{\rm bo}=10^{-11} ~ \rm ~ g ~ cm^{-3}$, and for  $n=3/2$, $\rho_{\rm bo}=10^{-9} \rm ~ g ~ cm^{-3}$ $\beta_{\rm bo}\leq0.02$, and   $n=3$, $\rho_{\rm bo}=10^{-9} \rm ~ g ~ cm^{-3}$ for all values of $\beta_{\rm bo}$.
In the same simulations we use a script of snapshots of the hydrodynamics for ($\rho,v)$ from a gray diffusion simulation started with the exact same initial conditions and resolution  \citep[see description of the numerics in][]{morag_shock_2023}. Convergence is achieved to the same extent in all cases.

\subsection{Planar SBO Temperature Results}
\label{sec: pSBO results}
\begin{figure}
\includegraphics[width=\columnwidth]{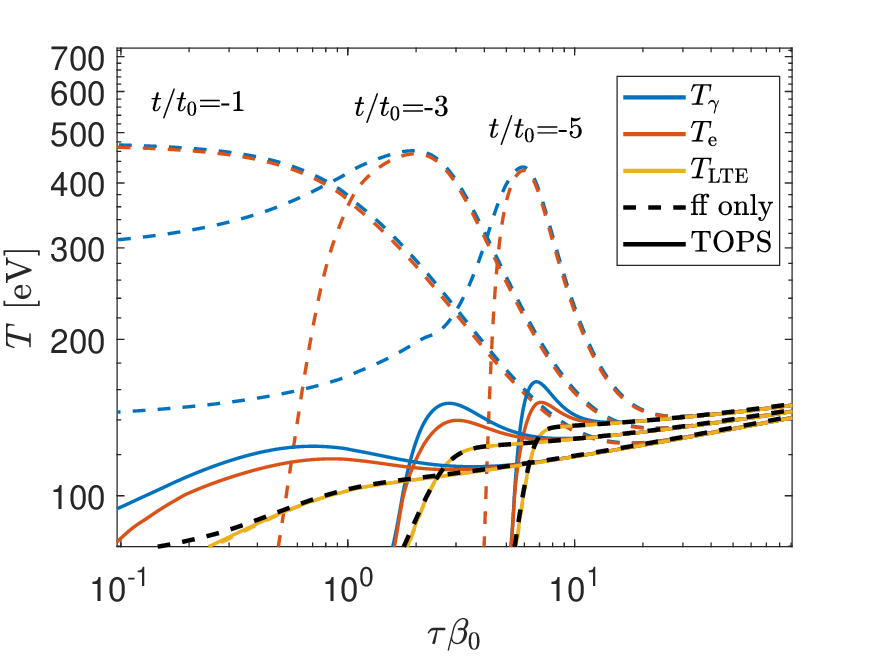}
  \caption{Snapshots of the temperature, 1 3 and 5 shock crossing times prior to shock breakout, for the choice $\rho_{\rm bo}=10^{-9} \, \rm g \, cm^{-3}$ and $\beta_{\rm bo}=0.1$. Peak temperature at the shock is much lower with TOPS opacity (solid lines - color) relative to ff-only (dashed lines -color). The equivalent LTE tempeature  (yellow lines) are compared to the Sakurai-Weaver Anzats (black dashed lines) showing excellent agreement. The legend only refers to the colored lines. The dashed black line is equation \ref{eq: T LTE S-W} based on the Sakurai-Weaver Anzats, added to show good agreement with the numerics.
  }
    \label{fig: pSBO T profile}
\end{figure}

We present numeric results for the planar shock-breakout problem, in two separate sets of simulations, (1) with free-free only emission, and (2) with TOPS opacity. The hydrodynamics and bolometric luminosity are insensitive to the choice of opacity. The hydrodynamics $(\rho,v,p)$ agree with the Sakurai-Weaver solution \citep{sapir_non-relativistic_2011}, to better than 2\% for all simulations\footnote{See also example agreement with $T_{\rm LTE}$, figure \ref{fig: pSBO T profile} prior to shock breakout.} for the times $t/t_{\rm bo}=-3,-5$ with no fitting of any kind.
At shock breakout at low velocities ($\beta\lesssim0.1$) we find good agreement ($3\%$ or better) between the bolometric luminosity in the simulations and the semi-analytic shock breakout description, given in $L_{SKW}$. For higher velocities, the bolometric luminosity can deviate from $L_{\rm SKW}$ by up to $15\%$, in agreement with \citetalias{sapir_numeric_2014}. The deviation is expected, and is due to the use of the Eddington approximation at the outer edge of the material (see discussion \citetalias{sapir_numeric_2014} in sec. 4.3).

\begin{figure}
\includegraphics[width=\columnwidth]{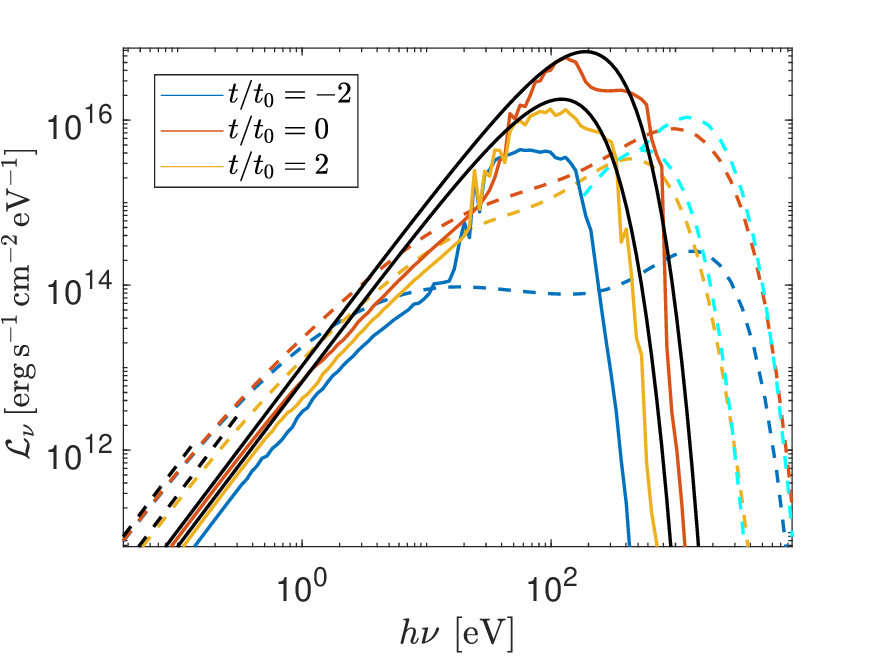}
  \caption{Output spectrum from the same simulations as figure \ref{fig: pSBO T profile}, shown near shock-breakout time. The spectrum at breakout in the case of free-free only opacity (dashed lines) is dominated by a Compton peak, in agreement with the semi-analytic description from \citetalias{sapir_non-relativistic_2013} (teal dashed lines). The output spectrum in the case of TOPS opacity (solid colored lines) is well described a Planck spectrum with $T_{\rm LTE,\mathcal{L}}$ (solid black lines - equation \ref{eq: T L LTE pSBO}).}
    \label{fig: pSBO Lnu}
\end{figure}

An example of the effects of opacity on planar shock breakout are shown in figures \ref{fig: pSBO T profile} and \ref{fig: pSBO Lnu} for the case $\beta_{\rm bo}=0.1$, $\rho_{\rm bo}=10^{-9} \, \rm g \, cm^{-3}$, exhibiting similar behavior to the uniform RMS problem. Figure \ref{fig: pSBO T profile} shows an example temperature profile. For Bremsstrahlung only (dashed lines), electron temperature and photon temperature deviate largely from $T_{\rm LTE}$ and are coupled in Compton equilibrium at the shock and immediately downstream of it. At shock breakout, $T$ will reach $\sim 500$ eV, in $\lesssim 10\%$ agreement with equation \ref{eq: T_peak SKWIII}. We observe a radiative precursor, similarly to the uniform RMS case and in agreement with \citetalias{sapir_numeric_2014}. Meanwhile, when TOPS opacity is included (solid lines), the temperature at peak remains much closer to LTE ($\sim150$ eV), and thermalizes more quickly in the downstream. During shock breakout, temperature upstream of the peak (towards the outer edge), decreases relative to the peak temperature.

Figure \ref{fig: pSBO Lnu} shows an example of the SED output spectrum near shock breakout, for the same simulations as in figure \ref{fig: pSBO T profile}. When Bremsstrahlung is used, photons above the critical frequency $h\nu_{\rm c}\sim 1 eV$ (see sec. \ref{sec: URMS intro}) up-scatter with the hot electrons to temperature $\sim3T$. The up-scattered photons dominate the spectrum, with typical temperature $T_\gamma$ and a Comptonized Wien spectrum, in agreement with the semi-analytic description of \citet{sapir_non-relativistic_2013} (hereafter \citetalias{sapir_non-relativistic_2013}, teal dashed lines). At frequencies below $h\nu_{\rm c}$, the photons are in thermal equillibrium with the plasma, forming a Planck spectrum with the same temperature $B_\nu(T)$ (black dashed lines). No fitting or rescaling is performed during the comparison to \citetalias{sapir_non-relativistic_2013}.

When TOPS opacity is included in figure \ref{fig: pSBO Lnu} (solid lines), the emission is similar to a Planck profile (example in solid black line for t=0). 10's of percent deviations from Planck are observed in the shape of the SED, primarily at intermediate frequencies ($h\nu \sim 10$ eV). At these frequencies, there are no atomic transition lines, and the emission opacity is relatively low. As a result, photons that would eventually escape the ejecta are produced deeper inside at larger optical depths, and their outward flux is suppressed by scattering. Outside of these frequencies there is larger opacity (free-free processes at lower frequencies and atomic transition lines at higher frequencies), and the spectrum is relatively close to Planck. At frequencies above the Planck peak, where atomic lines are generally strong, the emission is slightly reduced by line damping. These SED shape results are very similar to those of \citetalias{morag_shock_2024}, and we refer the reader to an in-depth discussion there.

\begin{figure}
\includegraphics[width=\columnwidth]{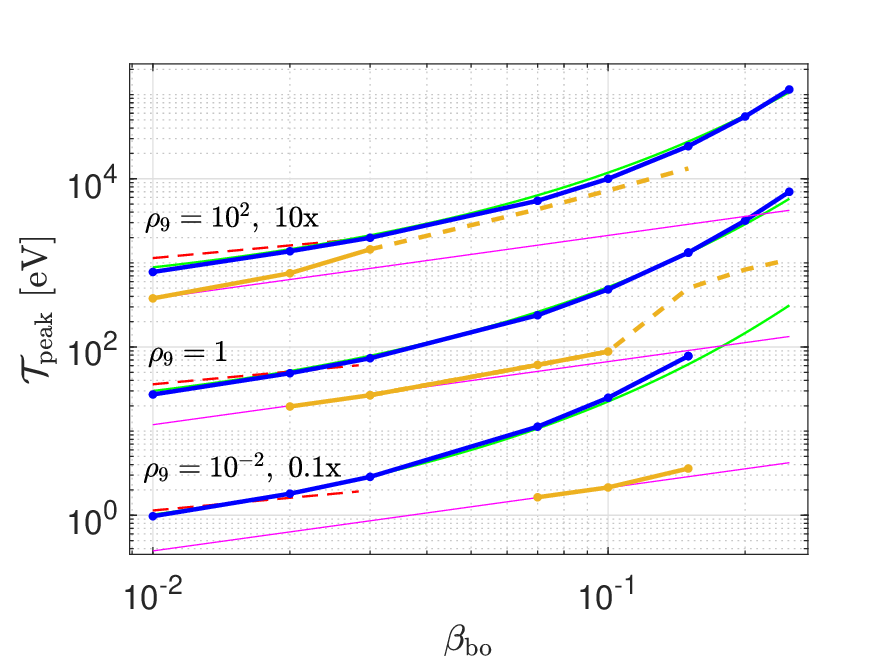}
\caption{
Mapping of peak emission temperature as a function of ($\rho_{\rm bo},\beta_{\rm bo})$ for density profile with $n=3/2$ (see text). Temperatures are visibly reduced in the presence of bound-free and bound-bound opacity due to reprocessing. We rescale the temperatures for lowest and highest densities for improved visibility. $\rho_9$ is defined by $\rho_{\rm bo}=10^{-9} \rho_9 ~ \rm ~ g ~ cm^{-3}$.
}
\label{fig: pSBO mapping n=1.5}
\end{figure}

\begin{figure}
 \includegraphics[width=\columnwidth]{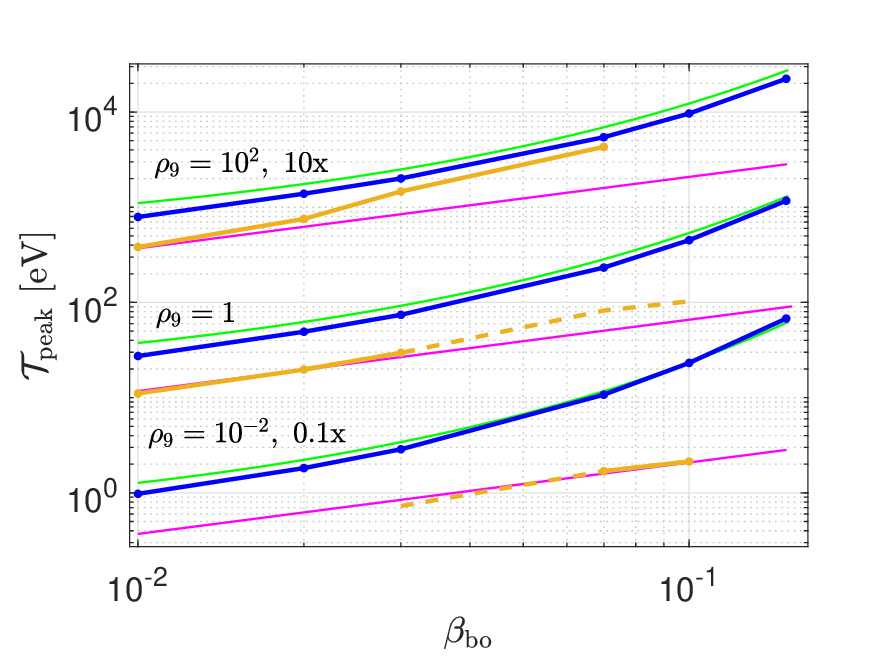}
\caption{
Same as figure \ref{fig: pSBO mapping n=1.5}, but with $n=3$.
}
\label{fig: pSBO mapping n=3}
\end{figure}

Figures \ref{fig: pSBO mapping n=1.5} and \ref{fig: pSBO mapping n=3} provide a mapping of the peak emitted breakout temperature as function of ($\rho_{\rm bo},\beta_{\rm bo}$), for $n=3/2$ and $n=3$, respectively. Simulation results in the case of free-free opacity (solid blue lines) are in good agreement with the analytic formula by \citet[][equation \ref{eq: T_peak SKWIII} - green lines]{sapir_non-relativistic_2013}, calculated assuming Compton equilibrium. At low velocities ($\beta\lesssim0.02$) in the presence of free-free opacity only, the radiation at the shock remains near LTE. Photons that escape during breakout are produced at $\tau \sim c/v_{\rm bo}$ (the immediate downstream), where the pressure is $p\sim \rho_{\rm bo} v_{\rm bo}^2$. Therefore, the peak emission temperature at these low velocities is approximately described to 10's of percent by eq. \ref{eq: T LTE RMS peak} that was introduced in the context the uniform radiation mediated shock, but with $\rho_{\rm bo}$ and $v_{\rm bo}$ replacing $\rho_0$ and $v_0$ (red dashed line).

The situation is different when bound-free and bound-bound processes are introduced (yellow lines). At slow velocities, photons produced at $\tau\sim\beta^{-1}$ reprocess on the way out towards the outer edge. Therefore, the emitted temperature represents the local temperature at $\tau \sim 1$, which itself is determined self-consistently by the outgoing flux. Therefore, the emission temperature at low velocities is well approximated as
\begin{equation}
    T_{\rm LTE,\mathcal{L}}=(4 \,\mathcal{L}_{\rm SKWI}\, / c\,  a_{\rm BB})^{1/4},
    \label{eq: T L LTE pSBO}
\end{equation}
where $\mathcal{L}_{\rm SKWI}(t)$ is the semi-analytic solution for the bolometric breakout luminosity, provided in \citet{sapir_non-relativistic_2011}. Then the peak luminosity (magenta line in figures \ref{fig: pSBO mapping n=1.5} and \ref{fig: pSBO mapping n=3}) is
\begin{equation}
    T_{\rm LTE,\mathcal{L} peak}=(4 \,(a_{\rm L} \rho_{\rm bo} v_{\rm bo}^3)\, / c\,  a_{\rm BB})^{1/4},
    \label{eq: T L LTE pSBO peak}
\end{equation}
where $a_{\rm L}=(0.72,0.77)$ for $n=3/2,3$ are the coefficients in \citet{sapir_non-relativistic_2011} representing the peak luminosity\footnote{As a sanity check, we also extract the equivalent temperature from $T_{\rm Lsim,LTE}=(4 \mathcal{L} / c\,  a_{\rm BB})^{1/4}$, where $\mathcal{L}$ is the bolometric luminosity extracted from the simulation. The result is the same as equation \ref{eq: T L LTE pSBO}}. Everywhere where equation \ref{eq: T L LTE pSBO} holds, and significant reprocessing of the photons occurs towards $\tau\sim1$, the observed emission temperature is at least a factor of two lower than the predicted temperature in the case of free-free only\footnote{For part of the parameter space, Compton equillibrium and deviation from LTE can be achieved at t=c/v during shock breakout, but emission can still be at LTE and agree with equation \ref{eq: T L LTE pSBO} due to reprocessing up to tau=1 (e.g. high velocities at low densities - $\rho_{\rm bo}=1e-11 ~\rm g~ cm^{-3}$, $\beta_{\rm bo}=0.1$).}. The opacity change also affects the time of peak temperature. For the free-free only case, $\mathcal{T}_{\rm peak}$ occurs roughly a shock crossing time prior to the time of peak luminosity \citep{sapir_non-relativistic_2013}, but with TOPS the $\mathcal{T}$ and $\mathcal{L}$ peaks coincide.

At higher velocities in the presence of TOPS, there is eventually a transition to a Comptonized (or partially Comptonized) output spectrum, tending towards the previous free-free temperature result (equation \ref{eq: T_peak SKWIII}). This transition implies both a reduced relative photon production rate at the peak due to high ionization, as well as limited reprocessing of the photons as they traverse from $\tau \sim c/v$ to $\tau\sim1$. 
At highest densities ($\rho_{\rm bo}=10^{-7} \rm ~g~cm^{-3}$), the transition to Compton behavior begins at lower velocities, $\beta_{\rm bo}\lesssim0.03$. At lowest densities ($\rho_{\rm bo}=10^{-11} \rm ~g~cm^{-3}$), the transition is barely observed at all, leading to a possible orders of magnitude difference in emission temperature when TOPS opacity is added.

Mapping the parameter space in the cases of bound-free and bound-bound opacity was not everywhere possible, partially due to numerical issues associated with the use of the TOPS table. In some cases, our simulations were converged to 10's of percents. In figures \ref{fig: pSBO mapping n=1.5} and \ref{fig: pSBO mapping n=3} we show some of these partially converged results in order to qualitatively demonstrate the transition towards `Comptonized' behavior (yellow dashed lines without dots). We also note that our temperature results at the transition are limited in their validity to a certain extent since our opacity assumes the plasma to be in thermal equilibrium with the radiation, which is not the case there.  Presumably at even higher temperatures, where the plasma is nearly fully ionized, this concern is again less important.

\section{Doppler Broadening and finite}
\label{sec: Doppler}
\begin{figure}
    \centering
    \includegraphics[width=\columnwidth]{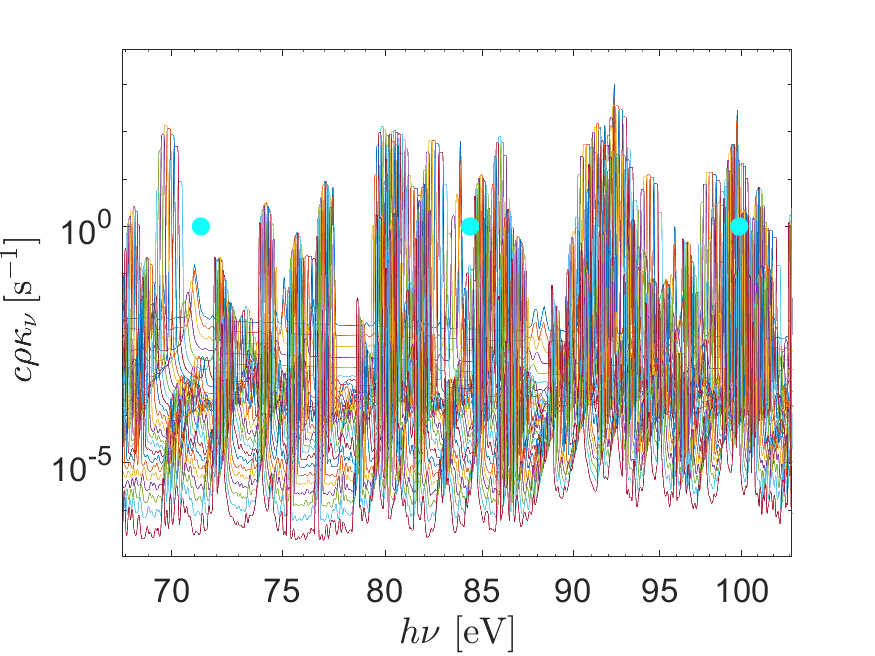}
    \caption{Example high-resolution TOPS opacity extracted using ($\rho,T$) from our simulation. Here ($n=3/2$, $\rho_{\rm bo}=10^{-9} \, \rm g \, cm^{-3}$, $\beta_{\rm bo}=0.1$). Plasma temperatures are in the range 60-100 eV, the time is t=0, corresponding with breakout in the equivalent Sakurai problem. Each grid location is Doppler shifted and smeared according to $v$ and $\Delta v$ from the same simulation. Only grid cells for $\tau<c/v$ are shown. The three large teal dots provide an example of the frequency edges of the binned multigroup photon groups.}
    \label{fig: rho kappa hi-res}
\end{figure}

Our code does not include expansion opacity and includes a finite frequency grid that is much coarser than the atomic line widths and the frequency separation between lines. As such, there is concern that strong lines can lead the simulation code to overestimate photon production (see in depth discussion in \citetalias{morag_shock_2024}). Similarly to \citetalias{morag_shock_2024}, we show here that Doppler broadening of the lines and the presence of a line forest in this problem, can smear the atomic line forest to an extent that greatly reduces the unevenness in $u_\nu$ across a frequency grid point, qualitative supporting the validity of our results. See an example in figure \ref{fig: rho kappa hi-res}, showing smearing of lines in $\rho\kappa_\nu$ across $\tau \sim c/v$ due to line expansion, where $\tau$ is the electron scattering opacity.

A second effect that is expected to smear $u_\nu$ during shock breakout is inelastic Compton scattering. Prominent lines in shock breakout temperatures can be relatively weak (e.g. $\kappa_\nu \sim O(1) ~ \rm cm^2 ~ g^{-1}$)\footnote{The situation is very different during shock cooling (e.g. $T\sim1$ eV), where inelastic Compton scattering is weak, and line strengths can reach $\kappa_\nu\sim O(10^6) ~ \rm cm^2 ~ g^{-1}$.}. Therefore, at breakout temperatures, the chance that a photon will escape resonance via Compton scattering with a cross section $\kappa_{\rm es}\sim0.34~ \rm cm^2 ~ g^{-1}$ becomes non-negligible. 

A single Compton scattering will shift the photon by $\Delta\nu/\nu \sim T/m_{\rm e}c^2$. For example, at $T=50$ eV, $\Delta\nu/\nu\sim10^{-4}$, a Doppler shift which is larger than the natural width of a line and comparable with the thermal line width $\Delta\nu/\nu\sim\sqrt{3 T / A m_{\rm p}}=5\times10^{-5}$, where we chose $A=55-65$. If a photon is scattered to a frequency and direction that will allow it to escape Doppler expansion opacity, it may escape resonance. Once outside of resonance, it should Compton scatter multiple times and efficiently diffuse in frequency.

In a previous work, we performed a relaxation rate analysis, concluding that LTE should be conserved up to recombination. Here, due to the relative higher density, we do not repeat this analysis and assume that the plasma degrees of freedom remain in LTE so long as the radiation does not deviate largely from Planck emission.

\section{Observational Impact}
\label{sec: Observations}

Here we estimate the effect that the introduction of atomic lines in the opacity will have on SN observations, including in X-ray. We estimate $\rho_{\rm bo}$ and $\beta_{\rm bo}$ for various types of progenitor stars using equations summarized in \citet{waxman_shock_2016}.
For red super giants:
\begin{align} \label{eq: rho_v_0_approx_RSG}
  \rho_{\rm bo} = 1.16 & \times 10^{-9} M_{0}^{0.32} {\rm v_{\ast,8.5}^{-0.68}} R_{13}^{-1.64} \kappa_{0.34}^{-0.68} f_{\rho}^{0.45}\, \rm g \, cm^{-3}, \nonumber\\
  {\rm v_{\rm bo}/v_{\ast}} & = 3.31 M_{0}^{0.13} {\rm v_{\ast, 8.5}}^{0.13} R_{13}^{-0.26} \kappa_{0.34}^{0.13} f_{\rho}^{-0.09}.
\end{align}
For blue super giants:
\begin{align} \label{eq: rho_v_0_approx_BSG}
  \rho_{\rm bo} = 3.1 & \times 10^{-10} M_{0}^{0.13} {\rm v_{\ast,8.5}^{-0.87}} R_{13}^{-1.26} \kappa_{0.34}^{-0.87} f_{\rho}^{0.29}\, \rm g \, cm^{-3}, \nonumber\\
  {\rm v_{\rm bo}/v_{\ast}} & = 4.1 M_{0}^{0.16} {\rm v_{\ast, 8.5}}^{0.16} R_{13}^{-0.32} \kappa_{0.34}^{0.16} f_{\rho}^{-0.05},
\end{align}
Based on these, we adopt the fiducial values given in \citetalias{sapir_non-relativistic_2013},
\begin{align}
    \beta_{\rm bo}&=0.05 M_{10}^{0.13} v_{*,8.5}^{1.13}R_{13}^{-0.26} \kappa_{0.4}^{0.13}f_\rho^{-0.09} \quad  \rm (RSG) \nonumber\\
    &=0.14 M_{10}^{0.16} v_{*,8.5}^{1.16}R_{12}^{-0.32} \kappa_{0.4}^{0.16}f_\rho^{-0.05} \quad  \rm (BSG) \label{eq: beta_bo RSG BSG WR}\\
    &=0.23 M_{5}^{0.16} v_{*,8.5}^{1.16}R_{11}^{-0.32} \kappa_{0.2}^{0.16}f_\rho^{-0.05} \quad  \rm (WR)  \nonumber
\end{align}
and
\begin{align}
    \rho_{\rm bo}&=2\times10^{-9}M_{10}^{0.32}v_{*,8.5}^{-0.68}R_{13}^{-1.64}\kappa_{0.4}^{-0.68}f_\rho^{0.45} ~ \rm g ~ cm^{-3} \,\, (RSG) \nonumber \\
&=7\times10^{-9}M_{10}^{0.13}v_{*,8.5}^{-0.87}R_{12}^{-1.64}\kappa_{0.4}^{-0.87}f_\rho^{0.45} ~ \rm g ~ cm^{-3} \, \, (BSG) \nonumber\\
&=2\times10^{-7}M_{5}^{0.13}v_{*,8.5}^{-0.87}R_{11}^{-1.26}\kappa_{0.2}^{-0.87}f_\rho^{0.29} ~ \rm g ~ cm^{-3} \, \, (WR), \label{eq: rho_bo RSG BSG WR}
\end{align}
where the progenitor radius is $R=10^x R_x$, and total mass is $M=x M_\odot M_x$, and $M_\odot$ is the solar mass.

Using equation \ref{eq: rho_v_0_approx_RSG}, we estimate the parameter range for RSG explosions, from a sample of 33 type II SNe in \citet[][]{irani_early_2023} - table 4. From light curve fits, they deduce progenitors in the range $R_{13}=3-18$, and $v_{\rm *,8.5}\approx0.3-2$\footnote{We employ the approximation $v_{s*}\approx 1.05 f_\rho^{-0.19}v_*\approx v_*$.}. This range of radii (in particular the lower values) is confirmed by the observed distribution of RSG's in the Magellenic clouds from \citet{davies_temperatures_2013} - see text in \citet{irani_early_2023}. As $f_{\rho}M$ is not discernible from early shock-curve fitting, we adopt $M=10M_\odot$, where $M_\odot$ is the solar mass, and $f_{\rho}\sim 1$, with minor effect on parameters.
We estimate therefore that explosions in RSG's primarily lie in the range $\rho_{\rm bo}\sim10^{-11}-3\times10^{-10} ~\rm g ~ cm^{-3}$, $\beta_{\rm bo}\sim0.01-0.05$, and therefore conclude that deviations from LTE for envelope breakout in RSG's are unlikely. This conclusion also holds for a more conservative value of $R_{13}=1$, i.e., the fiducial value $\beta_{\rm bo}=0.05$, $\rho_{\rm bo}=2\times10^{-9} ~ \rm g ~ cm^{-3}$ given in equations  \ref{eq: beta_bo RSG BSG WR} - \ref{eq: rho_bo RSG BSG WR}.

\begin{figure}
    \centering
    \includegraphics[width=\columnwidth]{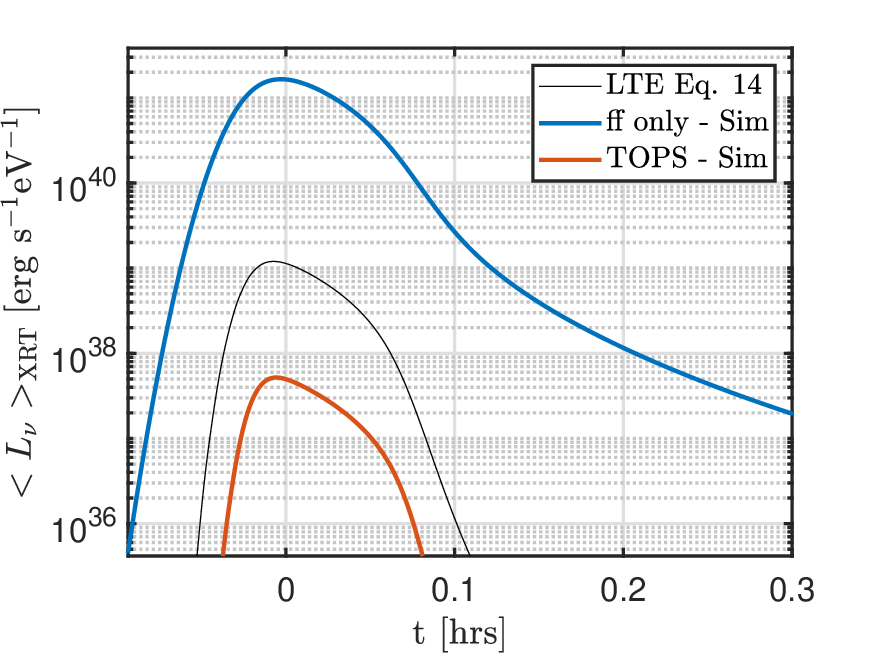}
    \caption{Demonstration of predicted X-ray shock breakout luminosities in the range [0.3,10] keV, representative of the Swift satellite XRT band and the Einstein Probe satelite. There is a $\sim3$ orders of magnitude difference in emitted energy between ff only and TOPS simulations. The approximate equation \ref{eq: T L LTE pSBO} agrees with the TOPS simulation up to an order of magnitude, as it does not include line dampening, which is important for part of the parameter space. This plot includes light travel time effects (equation \ref{eq: light travel time}). }
    \label{fig: Xray predictions}
\end{figure}

As a demonstration of the effect of temperature on type II SN observations, we show in figure \ref{fig: Xray predictions} light curve results in the Swift XRT band derived from SED's in our simulations. We choose an example $R=3\times10^{13}$ cm, $\rho_{\rm bo}=10^{-9} ~\rm g ~ cm^{-3}$, and $\beta_{\rm bo}=0.03$. The choice of radius enters into the luminosity description $L_\nu=4\pi R^2 \mathcal{L}_\nu$, where $\mathcal{L}_\nu$ is the planar SED from simulation. Radius also affects the emission due to light-travel time effects, which we include in the figure. These smear the observed emission in time by $t\sim R/c$, given by \citep{katz_non-relativistic_2012}
\begin{equation}
    L_{\nu, \, \rm obs}(t)=\int_0^1 h(\mu)L_\nu(t-R(1-\mu)/c)\mu d\mu,
    \label{eq: light travel time}
\end{equation}
where $h(\mu)=0.85+1.7\mu$.

In the presence of atomic lines in the above example, the peak equivalent photon temperature (equation \ref{eq: T_gamma}) is $\mathcal{T}_{\rm peak}=27$ eV, a factor of 2 lower than the case with free-free only opacity, $\mathcal{T}_{\rm peak}=73$ eV. Consequently, the flux in the XRT band ([0.3,10] keV), which lies in the Wien tail for both of these SED's, is lowered in the presence of TOPS by 3 orders of magnitude. In lower velocity and density simulations we find the X-ray flux, which is already exceedingly small $\sim10^{38} ~ \rm erg ~ s^{-1}$, can be reduced by even more orders of magnitude ($\sim10^{30} ~ \rm erg ~ s^{-1}$). Therefore, we conclude that the addition of atomic lines in the opacity reduces the volume of `envelope breakout' objects with observable X-Ray emission by at least $\sim(10^{-3})^{3/2}=3\times10^{-5}$ or even more (ignoring extinction), with sensitivity of this result to the low radius (higher $\beta_{\rm bo}$) distribution of RSG progenitors.

In blue super giants, the fiducial parameters for explosions based on equations \ref{eq: beta_bo RSG BSG WR} - \ref{eq: rho_bo RSG BSG WR}, are $\rho_{\rm bo}=7\times10^{-9} ~ \rm g ~cm^{-3}$ and $\beta_{\rm bo}=0.14$. SN 1987a, whose progenitor star exhibits $R\approx3\times10^{12}$ cm, $M\sim10 M_\odot$ \citep{arnett_supernova_1989} exhibits similar breakout parameters. Based on the results in figure \ref{fig: pSBO mapping n=3}, these values fall in the transition region, where the radiation distribution slightly deviates from local thermal equilibrium, but likely does not reach temperatures high enough to exhibit fully ionized Bermsstrahlung opacities (equation \ref{eq: T_peak SKWIII}). As blue super giant stars span a range, we expect some significant fraction of explosions from larger radius BSG's to remain in LTE following equation \ref{eq: T L LTE pSBO}, and the rest (at the fiducial value and above) to exhibit somewhat higher temperatures. Due to uncertainty regarding the TOPS opacities (and line transitions in general), the exact turn-off point in parameter space as well as the exact behavior in this intermediate region of the parameter space is difficult to predict.

Finally, Wolf-Reyet stars, which exhibit substantially smaller radii, and therefore significantly higher breakout densities and velocities ($10^{-7}~ \rm g ~cm^{-3}$, $\beta_{\rm bo}\gtrsim0.2$), will likely be fully ionized. Therefore, their observations should be unaffected by the the results of this paper.

\section{Discussion and Summary}
\label{sec: summary}

In this paper we numerically studied the uniform radiation mediated shock and the planar shock breakout problems, including for the first time, the effects of bound-free and bound-bound radiative processes. We found that for both systems, due to increased photon production, these processes can at times significantly delay deviations from local thermal equilibrium that were previously predicted for high velocities \citep[$\beta\gtrsim$ 0.07-0.1 - ][]{weaver_structure_1976,katz_fast_2010,sapir_non-relativistic_2013}.

In sec. \ref{sec: URMS intro}, we numerically solved the uniform RMS problem with both free-free only and TOPS (free-free, bound-free, bound-bound) opacity. In figure \ref{fig: U-RMS mapping}, we mapped the shock velocity and initial density parameter space ($\rho_0=10^{-11}-10^{-7} \rm ~ g ~ cm^{-3}$ and $\beta_0=c/v_0=0.01-0.25$).  In the free-free case, we observed temperature deviations from LTE towards Compton equillibrium, confirming previous analytic and semi-analytic estimates (equation \ref{eq: katz_Tpeak}). 
We found that for lowest densities, $\rho_0=10^{-11}~\rm g ~ cm^{-3}$, peak photon temperature when using TOPS opacities can remain in LTE (equation \ref{eq: T LTE U-RMS}) at high velocities. Meanwhile, at highest densities $\rho_0=10^{-7}~\rm g~cm^{-3}$, peak temperature is only mildly affected by the addition of bound-free and bound-bound opacity.

We perform a similar analysis of the planar shock breakout problem in \S~\ref{sec: pSBO}, including a mapping of peak temperature in figures \ref{fig: pSBO mapping n=1.5} and \ref{fig: pSBO mapping n=3} for initial density profiles $\rho\sim x^{3/2}$ and $x^3$, respectively. For free-free only simulations, peak temperature at shock-breakout is defined by the local temperature at an optical depth $\tau\sim c/v$. The observed peak Wien emission is well described to 10's of percent by equation \ref{eq: T_peak SKWIII} from the literature, which assumes Compton equillibrium, in agreement with the literature.

When TOPS opacity is employed in planar shock breakout, the production of additional photons can delay deviations from LTE. At emission temperatures $\mathcal{T}\lesssim100$ eV, photons get reprocessed as they escape. As we showed in figure \ref{fig: pSBO Lnu}, the emission is then will described by a blackbody $\mathcal{L}_\nu=\pi B_\nu$ with temperature given by equation \ref{eq: T L LTE pSBO} (our main analytic result - rewritten here)
\begin{equation}
    \mathcal{T}_{\rm LTE,\mathcal{L}}(t)=(4 \,\mathcal{L}_{\rm SKWI}\, / c\,  a_{\rm BB})^{1/4}.
    \label{eq: T L LTE pSBO 2}
\end{equation}
$\mathcal{L}_{\rm SKWI}$ is the \citet{sapir_non-relativistic_2011} bolometric breakout luminosity, provided as a table. Then the emission temperature at peak time is given by 
\begin{equation}
    \mathcal{T}_{\rm LTE,\mathcal{L} peak}=(4 \,(a_{\rm L} \rho_{\rm bo} v_{\rm bo}^3)\, / c\,  a_{\rm BB})^{1/4}.
    \label{eq: T Lpeak LTE pSBO 2}
\end{equation}
When LTE is maintained, these temperatures are at least a factor of 2 lower than the equivalent temperatures in the case of free-free only opacity.
At higher velocities and temperatures ($T\gtrsim 100$ eV), the system leaves LTE and transitions towards Compton equillilbrium even in the presence of atomic lines. In this instance the emission is determined again at $\tau\sim c/v$, and peak emission  temperature behaves according to equation \ref{eq: T_peak SKWIII}.

Our results affect the predictions for envelope shock breakout observations, especially with regards to x-ray (less so in the UV), shown in \S~\ref{sec: Observations}.
In figure \ref{fig: Xray predictions} we show the effect on X-rays.
For explosions in red super giants, strong deviations from LTE in breakout emission are unlikely. The deviation between our results and the previous literature prediction \citepalias{sapir_non-relativistic_2013} is roughly a factor of 2\footnote{\citepalias{sapir_non-relativistic_2013} also do not predict strong deviations from LTE due to the low densities and velocities for RSG shock breakouts}. Explosions from BSG's lie at the transition in behavior, such that for envelope breakout a fraction of them will remain near LTE, and a fraction may transition towards a Comptonized emission spectrum. Our results do not affect breakout from Wolf-Reyet stars, nor do they have an effect on predictions for GRB's associated with SBO. We also do not consider breakout in wind or circumstellar material which can lead to much different emission due to the presence of a  collisionless shock.

Velocity is an important predictor for the extent of deviations from LTE, especially in the case of free-free opacity. However, when bound-free and bound-bound processes are included, our mapping indicates that a primary factor that dictates the extent to whether $T_{\rm peak}$ will deviate from LTE is temperature (primarily the corresponding $T_{\rm LTE}$), due its effect on opacity. 


Our results were performed with the case of a solar mix of elements. For lower metallicity compositions, the photon production due to line emission may be reduced. We note however, that the shift away from LTE occurs primarily when the lines disappear altogether, such that we do not expect a slightly lower metallicity to have a pronounced effect on the result for much of the parameter space (similarly to \citetalias{morag_shock_2024}). Other systems where Hydrogen has been depleted (say in C/O dominated systems) the relative metallicity of heavy elements would only increase, along with the tendency to remain in LTE. A further study of the effect of metallicity on deviations from LTE is beyond our scope.

In \citet{morag_frequency_2023} we produced a frequency-dependent opacity table for arbitrary mixtures, based on the Kurucz atomic line list (see description in \citetalias{morag_shock_2024}). Kurucz contains atomic lines that are calibrated against experiment for temperatures near Hydrogen recombination  ($T\sim1\, \rm eV$). The most ionized states for heavy elements are not included. At breakout temperatures, where only the most ionized states (not fully ionized) from heavy elements contribute to the absorption opacity, the difference in opacity can be pronounced. In figure \ref{fig: knu_avg_example}, we provide an example of the multigroup opacity extracted from the hi-resolution opacity tables, TOPS and \citet{morag_frequency_2023}. At 40 eV, the averaged \citet{morag_frequency_2023} opacity is many orders of magnitude lower the corresponding TOPS opacity. This difference would lead to a much lower photon production rate.

\bibliography{references_local}

\end{document}